\documentclass{PoS}

\title{Limitations of cosmography in extended theories of gravity}

\ShortTitle{Limitations of cosmography in extended theories of gravity}

\author{\speaker{\'Alvaro de la Cruz-Dombriz}
\\
        Astrophysics, Cosmology and Gravity Centre (ACGC), Department of Mathematics and Applied Mathematics, University of Cape Town, Rondebosch 7701, Cape Town, South Africa \\
        E-mail: \email{alvaro.delacruzdombriz[at]uct.ac.za}}

\abstract{
The cosmographic approach, 
which only relies upon the homogeneity and isotropy of the Universe on large scales,
has become an essential tool in dealing with an increasing number of theoretical possibilities for explaining the late-time acceleration of the Universe, ranging from Modified Gravity theories to Dark Energy alternatives passing from testing the cosmological concordance $\Lambda$CDM model. Despite its generality, we show that this method 
has a number of shortcomings when trying to adequately reconstruct theories with higher-order derivatives in either the gravitational or the matter sector. Herein some paradigmatic examples of such an inability, explanations of the limitations and prospective cures will be presented.
          }

\FullConference{The 11th International Workshop Dark Side of the Universe 2015\\
		14-18 December 2015\\
		Kyoto, Japan}

\begin{document}

\section{Overview}
\label{S1}
Model-independent methods in cosmology have tried for years to make predictions on the {\it correct} underlying theory of gravity without making {\it  a priori } assumptions
on case-by-case classes of theories 
nor on specific models therein.  Such methods, which rely on some basic
 symmetries of the spacetime structure under consideration, have intended to infer the dark energy equation of state, reconstruct classes of dark energy theories and find any hints which unveil departures from both the Einsteinian gravity and the Copernican principle ({\it c.f.} \cite{Copeland:2006wr} for reviews). Among them, the assumption of the Copernican Principle (leading to the Friedmann-Lema\^{i}tre-Robertson-Walker (FLRW) metric) and the expression of the scale factor as a function of an auxiliary variable -- for instance either time or redshifts -- lead to the so-called  {\em cosmographic approach} \cite{Weinberg-Harrison} which aims to reconstruct the underlying cosmological dynamics using comparison to data derived from such two basic assumptions. {\it Cosmography} has thus tried to spot deviations from the standard $\Lambda$CDM model,  to reconstruct models for dark energy and to even shed some light on the form of the gravitational Lagrangian for classes of modified (extended) gravity theories \cite{Bernstein:2003es,Visser,Aviles_2012,Bamba:2012cp,Capozziello_PRD_fR,Capozziello_Ruth_PRD}.
 
 Technicalities of Cosmography are widely known and we refer the reader to more detailed literature \cite{WeinbergCosmography}. The most standard approach assumes as an
 auxiliary variable the usual redshift $z$ and then performs an expansion in the derivatives of the FLRW scale factor. Thus  the Hubble parameter and its derivatives can be 
 rewritten in terms of the  well-known cosmographic parameters.
One is then tempted to think that these parameters can be directly fitted with observational data, and that such a process leads to model (or theory)-independent constraints enabling reconstruction of the underlying cosmological models (or theories). As an alternative to the independent variable $z$, the above expansion may be expressed in terms of a variables ensuring the convergence of the series for the whole history of the Universe ({\it c.f.} \cite{Visser, DunsbyLuongo} for some possibilities). Nonetheless, 
the use of other redshifts, such as $y=z/(1+z)$ will be shown below to present strong limitations.

As a starting point, Cosmography was recently shown \cite{DombrizPRD2015} to suffer from shortcomings even to target $\Lambda$CDM as the underlying theory 
when SNIa mock data  \cite{union2.1} are precisely generated from a $\Lambda$CDM model and studied by means of the cosmographic method when several expansion orders or auxiliary redshift are considered. 
In Table \ref{tables1} as taken from \cite{DombrizPRD2015} it is shown how the $y$-parametrisation gives completely biased 
estimators, the trends being as follows: the $y$-parametrisation provides much bigger errors, biasing $q_0$ to smaller values and $j_0$ and $s_0$ to greater values than the true ones. In fact only a few a extensive simulations lie within $1\sigma$ whereas most of them do at 3 $\sigma$ or more \cite{DombrizPRD2015}. 
Moreover, the consideration of different number of parameters, namely ${\bf \theta_1}$ and ${\bf \theta_2}$ as described in Table \ref{tables1}, 
in the cosmographic expansion does not fix this limitation to correctly trace the exact values for the $\Lambda$CDM cosmographic parameters. 
In fact, as seen in Table \ref{tables1} as taken from \cite{DombrizPRD2015}, the inclusion of one extra parameter (namely $l_0$ in 
${\bf \theta_2}$) leads to error overestimates for both $z$ and $y$ variables. Consequently a {\it first caveat} in the standard cosmographic approach lies precisely in which variable to use: analysis in  \cite{DombrizPRD2015} led to conclude that despite the fact that $y$ might seem a more appealing variable from a theoretical point of view, it clearly turns out to not even be appropriate to derive cosmological constraints in $\Lambda$CDM, since fittings obtained in this variable are completely biased.

\begin{table}[htbp]
\caption{Coverage test for  two sets of parameters: ${\bf \theta_1} = \{H_0,q_0,j_0,s_0\}$ and ${\bf \theta_2} = \{H_0,q_0,j_0,s_0,l_0\}$. We refer the reader to \cite{DombrizPRD2015} for further details.}
\label{tables1}
\begin{center}
\begin{tabular}{@{}cccccccccccccc@{}}
\hline
&     &     &     &  ${\bf \theta_1}$   &      &     &     &      &     &  ${\bf \theta_2}$   &      &     \\ \hline 
& \vline    &   $y$   &   \vline  & \vline    &  $z$   &   \vline &   & \vline    &   $y$   &   \vline  & \vline    &   $z$   &   \vline  \\
\hline  & $1\sigma$ & $2\sigma$ &
 3$\sigma$ & $1\sigma$ & $2\sigma$ &
 3$\sigma$ &  & $1\sigma$ & $2\sigma$ &
 3$\sigma$ & $1\sigma$ & $2\sigma$ &
 3$\sigma$
\\ \hline
$q_0$ & 26 & 32 & 42 & 67   & 27 & 6 &  & 82 & 12 & 6 & 82   & 18 & 0  \\ 
$j_0$ & 10  &  45 & 45 & 64  & 29 & 7 &  & 93  &  5 & 2 & 88  & 12 & 0  \\ 
$s_0$ & 10  & 67 & 23 & 83  & 15 & 2 &  & 92  & 7 & 1 & 93  & 6 & 1 \\
$l_0$ & - & - & - & - & - & - &  & 100 & 0 & 0 & 100 & 0 & 0 \\
\hline
\end{tabular}
\end{center}
\end{table}

\begin{figure*}
\begin{center}
\includegraphics[width=0.45\textwidth]{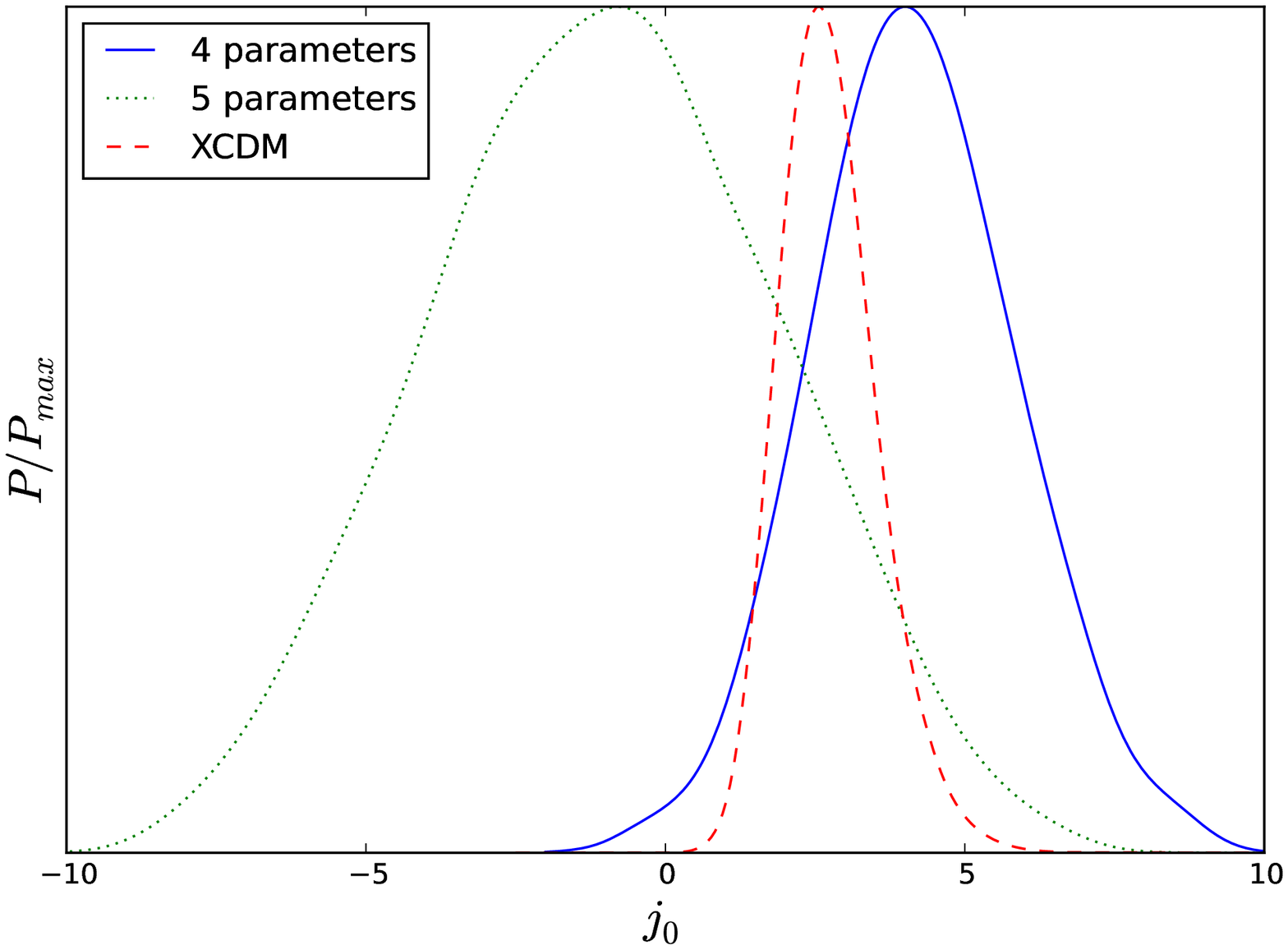}
\includegraphics[width=0.45\textwidth]{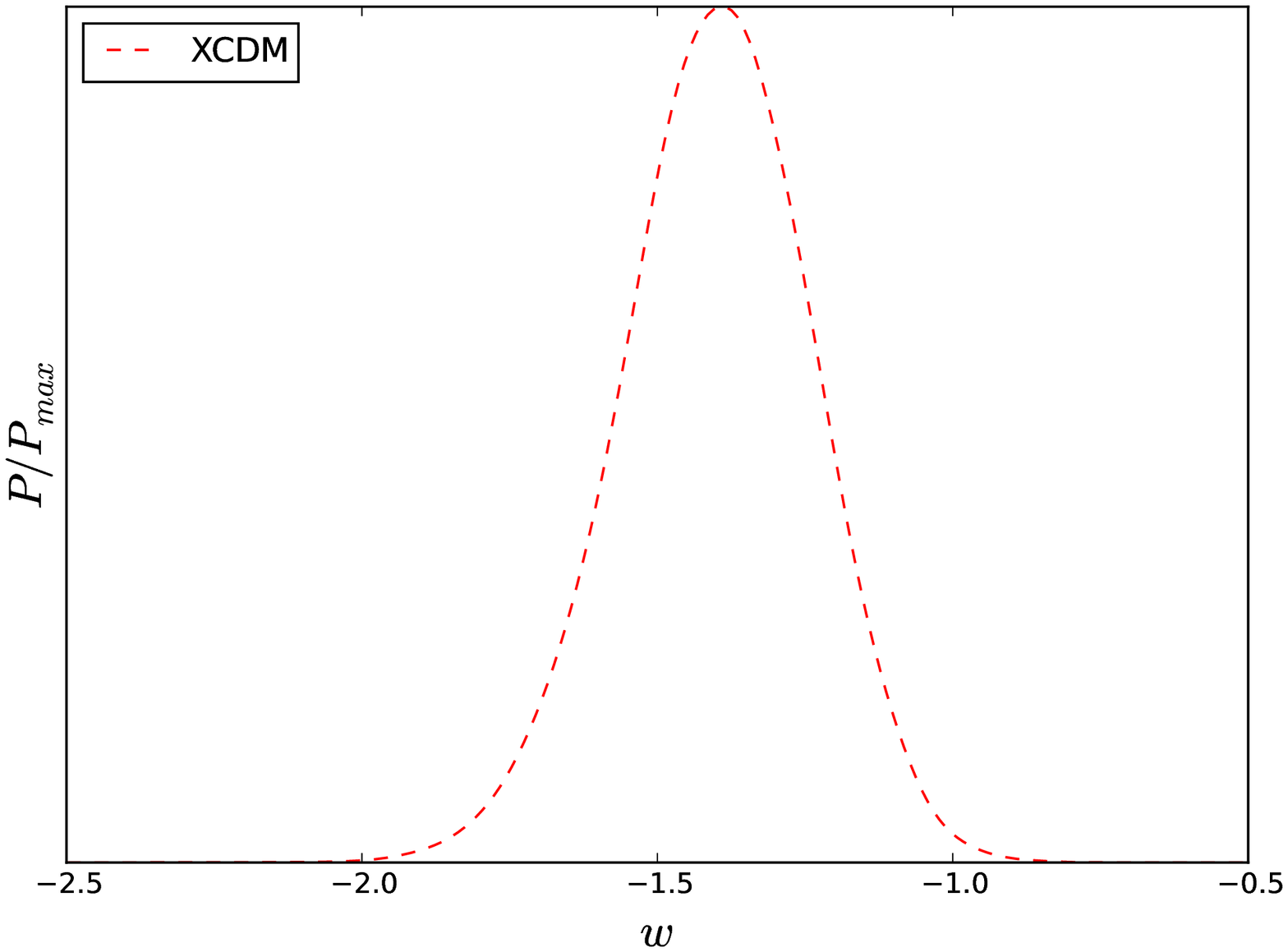}
\end{center}
\caption{As shown in  \cite{DombrizPRD2015}; Left panel: Posterior probability for $j_0$ considering cosmographic expansions in $z$-redshift of four parameters $({\bf\theta_1})$, five parameters $({\bf\theta_2})$ and the exact $X$CDM model $\{\Omega_m=0.3\,, \omega_X=-1.3\}$. Right panel: Posterior probability for the dark energy equation of state parameter $\omega$ in agreement with the $\omega=-1.3$ value.
}
\label{fig2}
\end{figure*}

Another consistency check 
of Cosmography consists of testing the ability of the method to spot the Concordance $\Lambda$CDM model against other competitive theories with constant (although slightly different from $\omega=-1$) dark-energy equation of state, i.e., $X$CDM models. Once again, the use of mock SNIa data can serve to illustrate the limitations of the method; this time those mock data have been obtained from a flat $X$CDM model with cosmological parameters $\Omega_m=0.3$ and $\omega_X=-1.3$. In Fig.~\ref{fig2} we have considered two
cosmographic realisations
 ${\bf \theta_1}$ (fourth order) and ${\bf \theta_2}$ (fifth order) in th  $z$-redshift expansion, which have then been 
compared to the direct fitting of free parameters $\{\Omega_m\,,\omega_X\}$ in the exact $X$CDM model. Results therein show that direct fitting to the $X$CDM model  provides smaller errors than the cosmographic approaches, so the direct fitting allows us to easily spot differences from $\Lambda$CDM model (for instance $j_0 \neq 1$), as expected from data which don't correspond to $\Lambda$CDM.
Moreover, the order of the expansion ultimately affects the posterior constraints on the cosmographic parameters: there is some evidence of $j_0 \neq 1$ 
for the  ${\bf \theta_1}$ set, which unexpectedly disappears for the ${\bf \theta_2}$ set. 

\section{Towards a better understanding of Cosmography in extended theories of gravity} 
The surprising limitations described in Sec.~\ref{S1} referred either to the drawbacks in the  Taylor expansions truncation or to the inability of targeting $\Lambda$CDM against simple dark energy theories, i.e., those with a minimal coupling and a constant equation of state representing the dark fluid. Nonetheless, the intense interest in the latest years on several types of gravitational theories beyond General Relativity, poses the question of the accuracy of cosmographic methods in spotting and eventually constraining such theories, or more to the point, of the intrinsic limitations that non-minimal couplings between geometry and matter and/or the presence of higher-order equations may have when such scenarios are subject to a cosmographic treatment. Most of the previous literature devoted to taking into account modern cosmography in the context of extended theories of gravity usually lacked any criticism and focused on finding constraints for classes of theories \cite{Capozziello_PRD_fR, Aviles_PRD_fT}. Moreover the literature at hand has often assumed General Relativity represents our Universe today exactly and thus assigned strict priors to the extra parameters. More recently \cite{DombrizPRD2015} has performed a more general treatment for both $K$-essence theories, scalar-tensor $f(R)$ theories and Galileons models, i.e., extended gravitational theories with either second order equations, higher-order equations or non-minimal couplings were addressed respectively there.
Difficulties in those three classes shared the common denominator of the presence of extra parameters, a fact that prevents us from finding a one-to-one correspondence between the derivatives of the theory and the cosmographic parameters. Let's summarise some of the main results of \cite{DombrizPRD2015} as follows,

\subsection{$K$-essence theories}
For a theory with an action - in appropriate units - given by
\begin{eqnarray}
\mathcal{S}=\int {\rm d}^4x^{}\sqrt{-g}\left[ \frac{1}{2}R 
 - \frac{1}{2} \omega (\phi)
\partial_{\mu} \phi \partial^{\mu }\phi -V(\phi )+\mathcal{L}_{m}\right]\ ,
\label{ST1}
\end{eqnarray}
where $\mathcal{L}_m$ is the matter Lagrangian,  $\omega(\phi)$ the factor that renormalises the scalar field $\phi$ and $V(\phi)$ 
the potential,  the mapping between cosmographic parameters and derivatives of the potential $V$ and the factor $\omega$ when  evaluated today only requires fixing one extra parameter \cite{DombrizPRD2015}. Thus, the derivatives of the potential\footnote{Analogous results are obtained for the expansion of $\omega(\phi)$ derivatives evaluated today.} evaluated at $z=0$,  can be written\footnote{Field equations in a spatially-flat FLRW were used.} as
\begin{eqnarray}
&\frac{V_0}{H_0^2}&=2-q_0-\frac{3\Omega_m}{2}\ ,\;\;\; 
\frac{V_{z0}}{H_0^2}=4+3q_0-j_0-\frac{9\Omega_m}{2}\ , \nonumber\\
&\frac{V_{2z0}}{H_0^2}&=4+8q_0+j_0(4+q_0)+s_0-9\Omega_m\ ,\\
&\frac{V_{3z0}}{H_0^2}&=j_0^2-l_0-q_0j_0(7+3q_0)-s_0(7+3q_0)-9\Omega_m\ .\nonumber
\label{ST7}
\end{eqnarray}
The standard assumption herein in order to recover a one-to-one correspondence consists of connecting $\Omega_m$ and the cosmographic deceleration parameter $q_0$ as in the $\Lambda$CDM model, i.e., 
%
$\Omega_m \approx 2/3(1+q_0)$. Despite this assumption, cosmography can be shown \cite{DombrizPRD2015} to perform far worse than other model-independent approaches such as Gaussian process regression \cite{nair} for which errors are much smaller and no assumption about the model behaviour today needs be made, so any attempt at reconstructing $K$-essence theories following a cosmographic approach seems to be disfavoured with respect to Gaussian regression. Moreover, dependence in the order expansion is also present, resulting with after $z \sim 0.5$ no useful constraints for $V(\phi)$ are derived for a ${\bf \theta_2}$ realisation, whereas for the ${\bf \theta_1}$ counterpart good constraints are obtained up to $z \sim 1$, showing the fragility of the method for these theories.

\subsection{$f(R)$ theories}

The understanding of limitations in cosmography for theories involving higher-order equations, can be exemplified by the 
paradigmatic $f(R)$ scalar-tensor theories which in the metric formalism give rise to fourth-order equations. Once again, a mapping between cosmographic parameters and derivatives of the gravitational Lagrangian $f(R)$ can be found \cite{DombrizPRD2015}. The price to pay is the existence of  two extra free parameters which can be thought of as the first and the second derivatives of the $f(R)$ Lagrangian evaluated today.
Previous works \cite{Capozziello_PRD_fR} fixed those values so that the $f(R)$ Lagrangian was forced to coincide with General Relativity at $z=0$, namely $f_R(z=0)=1$ and $f_{RR}(z=0)=0$. However such naive priors may lead to 
either singularities or instabilities occuring \cite{Pogosian:2007sw}, apart from the fact that cosmological values for $f(R)$ derivatives today may be different from GR exact values and still produce viable cosmological models \cite{SulonaPRD2016}. Thus one is led to 
abandon the  one-to-one correspondence between the $f(R)$-derivatives and the cosmographic parameters, so 
either priors over these $f(R)$ parameters or alternative tests are 
required \cite{Refs_fR_tests}. 
Whenever either those priors are too strict or marginalisation of extra parameters is not carefully performed, the obtained constraints 
can lead to a reconstruction of the  $f(R)$ models 
which are not capable of generating the SNIa mock data used to test the method.
A combination a wide set of priors through extensive Markov Monte Carlo Chain  simulations is performed in \cite{Reverberi_Cosmo_2016} and is expected to improve the predictability of the method also making use of other astrophysical probes, such as Baryon Acoustic Oscillations and $H(z)$ measurements.

\subsection{Galileons theories}
Finally, we could exemply the cosmography limitations when dealing with theories which have higher-order derivatives but this time in the matter sector, such as Galileons 
\cite{Deffayet:2009mn}. Thus for one of the simplest Galileon models involving three coupling constants, analogous issues as those described for $f(R)$theories show up, due to the existence of two additional free parameters. As remarked in \cite{DombrizPRD2015}, 
 the sole advantage of cosmographic treatment for Galileons theories is that since the gravitational sector does not involve higher-order derivatives the errors are not as large as those obtained in the  $f(R)$ gravity case.  This could also be understood by the fact that  the expansion of the  scalar field for Galileons up to second order only depends on $q_0$ and not on higher cosmographic parameters \cite{DombrizPRD2015}.

\section{Conclusions and Prospects}
At the present stage, the cosmographic approach seems to be plagued with drawbacks, previously overlooked in the literature, such as the dependence of results with the chosen independent variable in which the cosmographic expansion is performed as well as the lack of hints of the required expansion order which minimises the induced errors. These two facts can for instance lead to not correctly spotting fiducial models which slightly differ from the Concordance cosmological $\Lambda$CDM model.

Moreover, the reconstruction process based on cosmography for theories containing extra degrees of freedom, such as quintessence-like theories, $f(R)$ gravities or Galileons and, by extension  other modified theories, seem to lead to completely unconstrained parameters of the models under consideration when either $a)$ Einsteinian gravity is assumed as a benchmark for models' behaviour today, or $b)$ narrow priors or limited marginalisations for extra parameters are considered. Such cosmographic treatment may prevent us from ruling models which exhibit very 
unrealistic cosmological background evolutions and more dangerously, could lead to think of the viable character of models that 
are already excluded thanks to the use of other methods. %
 
 Consequently, it might seem that the cosmographic approach  - as traditionally thought - is a strongly limited tool for theory reconstruction. In fact other limitations, apart from those explained in the bulk of this communication might appear, namely there are at least two effects which, for the sake of simplicity, were neglected in this communication. 
The first one considers the eventual role of the spatial curvature $\Omega_k$ in fixing constraints, due to the fact that $\Omega_k \neq 0$ can induce a time variation for the dark energy equation of state \cite{chris2007} and eventually distort the cosmological constraints as recently claimed in \cite{Leonard:2016evk}.
The second, additional limitations concerns other effects such as both gravitational and Doppler lensing \cite{grav_lens}, or even local gravitational redshifts \cite{wojtak}, which may involve the appearance of extra scatters in the Hubble diagram, a fact which might degrade cosmological constraints as obtained from SNe Ia data analysis. Although quantified and included as an extra error
SNe Ia analysis within the paradigm of  the $\Lambda$CDM model  \cite{union2.1}, the impact that such scatters may have in theories of modified gravity, still requires  a thorough study to be able to determine for which extended theories the cosmographic approach is expected not to be useful in constraining parameters.

But not everything is negative: as possible routes which deserve further exploration with the hope of curing limitations in Cosmography we could mention:
\begin{enumerate}
\item the need for a clear definition of the adequate auxiliary variable(s), together with the comprehension of their range of validity and extensive testing against mock data from several catalogues. In this respect, the determination of a 
convenient {\it pivot redshift}, around which the performed expansions become optimal could shed light on whether or not the lack of competitiveness of the cosmographic method remains. Clearly, such a {\it pivot} choice is expected to severely depend upon the redshift distribution of the considered astrophysical catalogue. In other words, this adaptive study may help us to understand how the redshift distribution of different samples influences the results of some ongoing and future surveys, such as DES and LSST.

\item  a robust statistical method involving, for instance, full freedom in the priors and a combined analysis of several probes which could in principle establish a correlation between the number of data points, location in the redshift space and the optimal number of cosmographic parameters, as well as Bayesian evidence enabling us to provide some criteria to rule out regions in the parameter space of classes of models.

\item Finally, as mentioned in the bulk of the communication, the use of well posed priors over the extra parameters in higher-order theories which allow us to obtain competitive constraints for such theories, even when lensing and local effects are neglected. Work in progress in \cite{Reverberi_Cosmo_2016} is in this direction for several classes of theories, namely scalar-tensor and extended theories of torsion.

\end{enumerate}

Finally, the aim of the present communication has been to 
illustrate the limitations of Cosmography by making sole use of  one observable, namely supernovae of type Ia, with the aim of diverting from the technicalities of a multi-observable treatment,
which might have obscured the origin of our results. Although mainly in the context of Einstein gravity different observables have been used to constrain cosmographic parameters, such as $H(z)$ data, baryon acoustic oscillations, gamma-ray bursts, angular distances to galaxy clusters \cite{cosm_constraints}, 
it is easier to understand both the cause and the extent of possible limitations with only one observable. We shall present a combined probes analyses, namely H(z), BAO and extend the supernovae treatment in our future work \cite{Reverberi_Cosmo_2016}.

\section*{Acknowledgments}
The author acknowledges financial support from the University of Cape Town (UCT) Launching Grants programme, National Research Foundation grant 99077 2016-2018, Ref. No. CSUR 150628121624, MINECO (Spain) projects FIS2014-52837-P, FPA2014-53375-C2-1-P,  CSIC I-LINK1019 and Consolider-Ingenio MULTIDARK CSD2009-00064.  He would also like to thank the Yukawa Institute for Theoretical Physics (Kyoto, Japan) for their hospitality and the invitation to present these results.

\end{document}